\documentclass[mypaper,7pt,twoside]{CoAst}
\usepackage{epsf,graphicx,fancyhdr}
\renewcommand{\chaptermark}[1]{\markboth{#1}{}}

\newcommand{\aap}{A\&A}  %*************Stimmts?
\newcommand{\mnras}{MNRAS}

\newcommand{\ngc}[1]{NGC~{#1}}                              % Style for NGCxxxx
\newcommand{\halpha}{H$\alpha$}  % H alpha

\def\cd {d$^{-1}$}

\newcommand{\kopf}{\small\itshape Comm. in Asteroseismology \\ Contribution to the Proceedings of the 38$^{th}$\,LIAC\,/\,HELAS-ESTA\,/\,BAG, 2008
}

\newcommand{\Authors}[1]{\begin{center}\normalsize\bf\sf #1 \end{center}}

\renewcommand{\author}[1]{\begin{center}\normalsize\bf\sf #1 \end{center}}
\newcommand{\Address}[1]{\begin{center}\small\sf #1 \end{center}}

\DeclareGraphicsExtensions{.eps,.jpg}

\newcommand{\Session}[1]{{\vspace{3mm}\small \noindent  \hspace*{3mm} Session: } #1 \normalsize}

\newcommand{\Objects}[1]{{\vspace{0mm}\small \noindent  \hspace*{3mm} Individual Objects: } \small #1 \normalsize}

	\newcommand{\four}{\small Observed frequencies in pulsating massive stars}

\renewenvironment{abstract}{\section*{Abstract}\normalsize\sf}{}
\newcommand{\References}[1]{\begin{flushleft}{\large References\\}\vspace*{2mm}\small #1 \end{flushleft}}

\newcommand{\chapterCoAst}[2]{\chapter[\sf\normalsize #1\\ \footnotesize \hspace*{5mm}by #2 \sf\normalsize][]{#1\\}\rhead[\fancyplain{}{\sf\footnotesize \center{#1}}]{\fancyplain{}{\sffamily\thepage}}\lhead[\fancyplain{\kopf}{\sffamily\thepage}]{\fancyplain{\kopf}{\sf\footnotesize \center{#2}}}}

\newcommand{\figureCoAst}[5]{\begin{figure}[#4]
\centering
\includegraphics*[#5]{#1}
\caption{#2}
\label{#3}
\end{figure}}

\renewcommand{\chaptername}{}

\newcommand{\acknowledgments}[1]{\vspace*{5mm}\noindent  \textbf{Acknowledgments.} #1}

\def\rfr{\smallskip\par\noindent
        \hangindent=7truemm
        \hangafter=1}

\begin{document}
\sf

\chapterCoAst{Asteroseismology of massive stars in the young open cluster \ngc{884}: a status report}%paper titel and page heading for even pages
{S.\,Saesen, F.\,Carrier, A.\,Pigulski, et al.} %page heading for odd pages
\Authors{S.\,Saesen$^1$, F.\,Carrier$^1$, A.\,Pigulski$^2$, C.\,Aerts$^1$,
G.\,Handler$^3$, A.\,Narwid$^2$, J.\,N.\,Fu$^4$, C.\,Zhang$^4$,
X.\,J.\,Jiang$^5$, G.\,Kopacki$^2$, J.\,Vanautgaerden$^1$,
M.\,St\c{e}\'{s}licki$^2$,
B.\,Acke$^1$, E.\,Poretti$^6$, K.\,Uytterhoeven$^6$, W.\,De\,Meester$^1$, M.\,D.\,Reed$^7$,
Z.\,Ko{\l}aczkowski$^2$, G.\,Michalska$^2$, E.\,Schmidt$^3$, R.\,\O{}stensen$^1$,
C.\,Gielen$^1$, K.\,Yakut$^{8,9}$, A.\,Leitner$^3$, B.\,Kalomeni$^{10}$, S.\,Prins$^1$,
V.\,Van\,Helshoecht$^1$, W.\,Zima$^1$, R.\,Huygen$^1$, B.\,Vandenbussche$^1$, P.\,Lenz$^3$,
D.\,Ladjal$^1$, E.\,Puga\,Antol\'{\i}n$^1$, T.\,Verhoelst$^1$,
J.\,De\,Ridder$^1$, P.\,Niarchos$^{11}$,
A.\,Liakos$^{11}$, D.\,Lorenz$^3$, S.\,Dehaes$^1$, M.\,Reyniers$^1$, G.\,Davignon$^1$,
S.-L.\,Kim$^{12}$, D.\,H.\,Kim$^{12}$, Y.-J.\,Lee$^{12}$, C.-U.\,Lee$^{12}$, J.-H.\,Kwon$^{12}$,
E.\,Broeders$^1$, H.\,Van\,Winckel$^1$, E.\,Vanhollebeke$^1$, G.\,Raskin$^1$, Y.\,Blom$^1$,
J.\,R.\,Eggen$^7$, P.\,Beck$^3$, J.\,Puschnig$^3$, L.\,Schmitzberger$^3$,
G.\,A.\,Gelven$^7$,
B.\,Steininger$^3$, and R.\,Drummond$^1$} 
\Address{
$^1$ Instituut voor Sterrenkunde, Katholieke Universiteit Leuven, Belgium\\
$^2$ Instytut Astronomiczny, Uniwersytet Wroc{\l}awski, Poland\\
$^3$ Institut f\"{u}r Astronomie, Universit\"{a}t Wien, Austria\\
$^4$ Beijing Normal University, China\\
$^5$ National Astronomical Observatories, Chinese Academy of Sciences,Beijing, China\\
$^6$ INAF-Osservatorio Astronomico di Brera, Merate, Italy\\
$^7$ Department of Physics Astronomy and Material Science, \\Missouri State University, USA\\
$^8$ Institute of Astronomy, University of Cambridge, UK\\
$^9$ Department of Astronomy \& Space Sciences, University of Ege, Izmir, Turkey\\
$^{10}$ Izmir Institute of Technology, Department of Physics, Izmir, Turkey\\
$^{11}$ Department of Astrophysics, Astronomy and Mechanics, \\University of Athens, Greece\\
$^{12}$ Korea Astronomy and Space Science Institute, Daejeon, South Korea\\
}

\noindent
\begin{abstract}
To improve our comprehension of the $\beta$\,Cephei stars, we set up a photometric multi-site campaign on the open cluster \ngc{884} ($\chi$ Persei). Thirteen telescopes joined the 2005-2007 campaign which resulted in almost 78\,000 CCD frames. We present an up-to-date status of the analysis of these data, in which several interesting oscillating stars are pointed out. We end with the future prospects.
\end{abstract}

\Session{\four}\\ % you can chose from, \one, \two, ... \six, \ESTA, \future
%or \poster
\Objects{\ngc{884}} 

\section{Introduction}
Recent progress in the seismic interpretation of selected
$\beta$\,Cephei stars was remarkable in the sense that standard
stellar structure models are unable to explain the oscillation data
for the best-studied stars: HD\,129929 (Aerts et al.~2003), $\nu$\,Eridani (Pamyatnykh et al.~2004, 
Ausseloos et al.~2004, Dziembowski \& Pamyatnykh~2008) and 12\,Lacertae (Ausseloos~2005, Handler et al.~2006, Dziembowski \& Pamyatnykh~2008). Non-rigid internal rotation and core convective overshoot are needed to fit the
measured oscillation frequencies and the standard models have now been
upgraded to include these effects, albeit in a crude parametrised way.
Pamyatnykh et al.~(2004) have suggested to include in future models radiative diffusion processes as well, in an attempt to resolve the yet unsolved excitation problem encountered for some of
the modes detected in $\nu$\,Eridani. 

A next step in asteroseismology of $\beta$\,Cephei stars was
undertaken recently, with the study of these stars in clusters. Indeed, with the current CCD cameras we are able to obtain simultaneous measurements of thousands of stars. Another big advantage is the cluster membership of the stars: this gives us much tighter constraints when
modelling their observed and identified oscillation modes.

Krzesi\'{n}ski \& Pigulski (1997, 2000) discovered one candidate and two bona fida $\beta$\,Cephei stars in \ngc{884}. The variability study on this cluster conducted by Waelkens et al.~(1990) showed that at least half of the brighter stars are variable, while most of them seem to be Be stars.

\section{Observations}
To perform ensemble-asteroseismolgy of \ngc{884}, we needed time-resolved multi-colour differential photometry of a selected field of this cluster, for which we organised a large-scale multi-site campaign. An international team moni-tored the cluster with 13~telescopes in the northern hemisphere in the filters U, B, V, I. The data were taken between 2005 and 2007, spread over three observational seasons, spanning in total 800~days. 77\,900~CCD images and 92~hours of photo-electric measurements were collected, representing 1290~hours of data taken by about 60~observers. A world map indicating all participating sites and their characteristics (telescope diameter and filters used) and the distribution of the observations in time can be found in Saesen et al.~(2008).

The effect of the multi-site character of the campaign is best seen when
comparing the spectral windows of the data (see Fig.~\ref{sw}). The spectral
window of the Polish data, the site that has the largest contribution to the whole data set,
has a high one-day-alias: its amplitude is 90\% of the main frequency peak. If
we add one site, China, situated at a different latitude, the
one-day-alias falls down to 70\% of the central peak. Taking all sites into
account makes the alias drop to only 55\%, which makes it much easier to
identify the correct frequency peak.
\figureCoAst{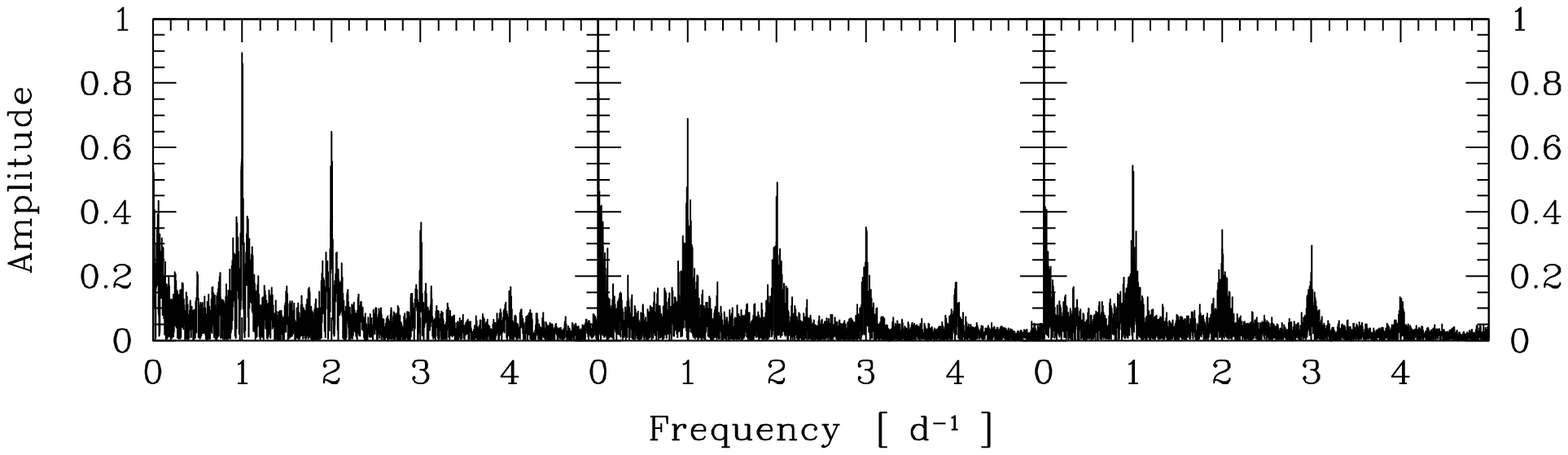}{Spectral windows of the
Polish data (left panel), the combined Polish and
Chinese data (middle panel) and all data (right
panel).}{sw}{!ht}{clip,angle=0,width=\textwidth}

To transform the CCD frames into interpretable light curves, we extracted the
fluxes of the stars with Daophot (Stetson~1987), in which we combined PSF and
aperture photometry. We performed differential photometry in which we correct for atmospheric extinction by taking several  reference stars distributed over the CCD frame into account. Currently we are detrending the data with SysRem (Tamuz et al.~2005) to remove the
linear systematic effects which are present in a lot of stars. For sites with many data points we
obtain an overall V~accuracy of 3-5\,mmag over the entire campaign, depending on the telescope. The error
on the frequency is smaller than 0.000\,14\,\cd\,and the detection threshold at
high frequencies for Polish data is about 0.3\,mmag. Frequencies in the
$\beta$\,Cephei range can be accepted if their amplitudes are larger than
1\,mmag, but this limit will go down once all data are put together.

\section{Detected variable stars}
A preliminary analysis on single-site data led to the confirmation of the two
known $\beta$\,Cephei stars, Oo~2246 and Oo~2299, and to the discovery of
numerous new pulsators of this type, among which several are multi-periodic
(Pigulski et al.~2007, Saesen et al.~2008). However, spectroscopy has shed
additional light on two of these newly classified $\beta$\,Cephei stars, Oo\,2085
and Oo\,2566. They show variations on both short ($\sim$hours) and long ($\sim$days) time scales and
turn out to have \halpha~emission. As a consequence,  Oo\,2085
and Oo\,2566 are also categorised as
Be stars. Oo~2444, Oo~2488 and Oo~2572 remain accepted as $\beta$\,Cephei stars.

The $\beta$\,Cephei stars show at least a double-mode behaviour, except Oo~2299, which seems to show a single, predominant mode. Fig.~\ref{betacep} shows the periodograms in subsequent
stages of prewhitening based on dual-site data of two of these oscillators. We
clearly detect two frequencies for each of them (first two panels) and after
subtracting these variations from the data, there is still power
present in the residuals (lower panels). We expect that additional frequencies
will peak above the frequency acceptance level after detrending and merging the
data of all sites. The case of Oo~2572 also points out that multi-site
observations are important to pinpoint the correct frequency peak. In Saesen et
al.~(2008), which is only based on single-site data, an alias frequency was
mistaken for the correct peak. 
\figureCoAst{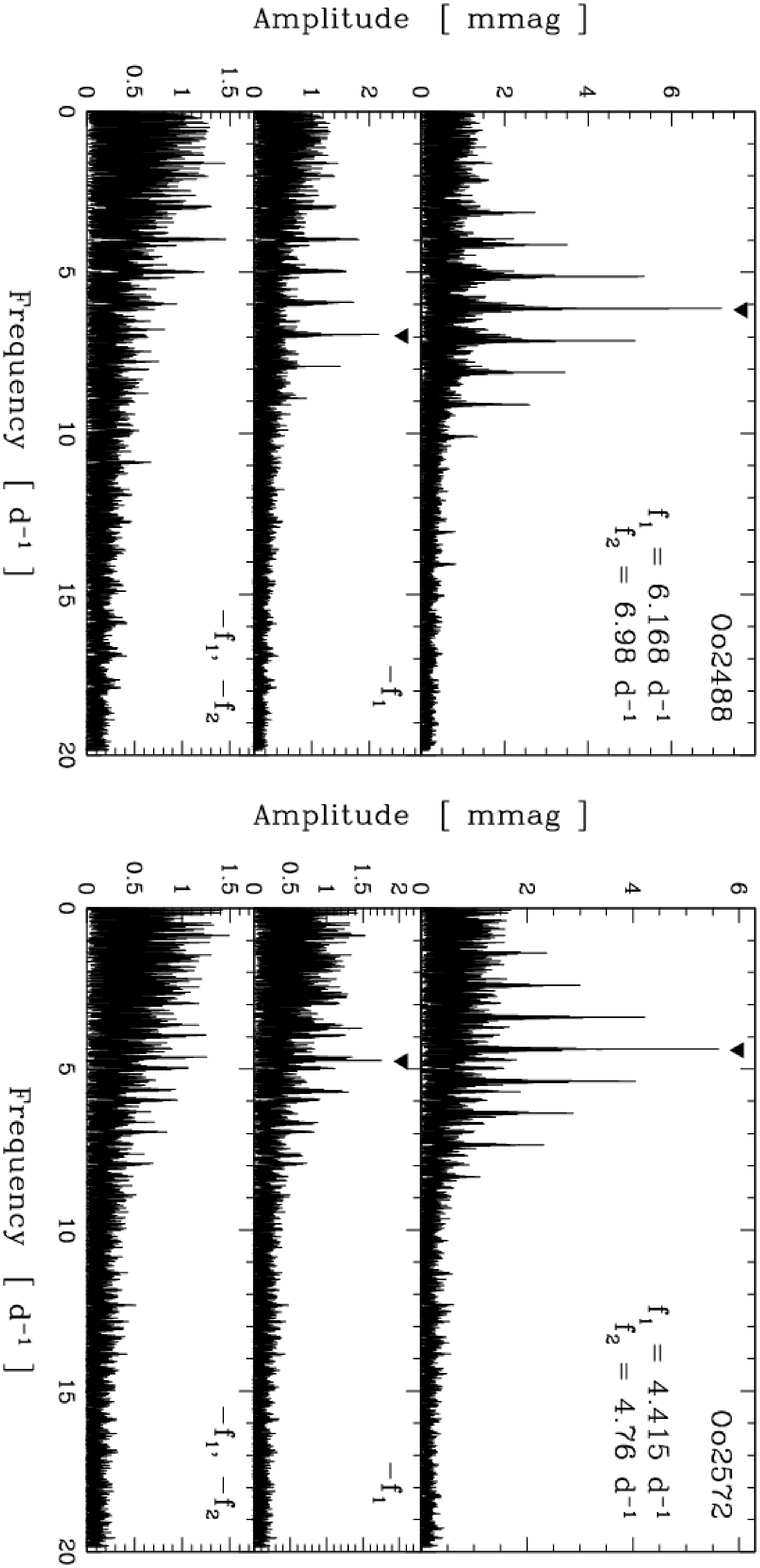}{Periodograms of subsequent stages
of prewhitening based on Polish and Chinese data of the
$\beta$\,Cephei stars Oo~2488 and
Oo~2572.}{betacep}{!ht}{clip,angle=90,width=\textwidth }

Besides these five established $\beta$\,Cephei stars, we have five more
candidates. Three of them show evidence
for low frequencies and might turn out to be hybrid oscillators. We have
detected seven SPB candidates, amongst which is Oo~2253: three significant
frequencies are extracted and shown in the subsequent periodograms of
Fig.~\ref{spb}. In addition,
the observed field of \ngc{884} contains several eclipsing binaries (see
Fig.~\ref{ecl} for two newly discovered cases) and other variable stars. 
\figureCoAst{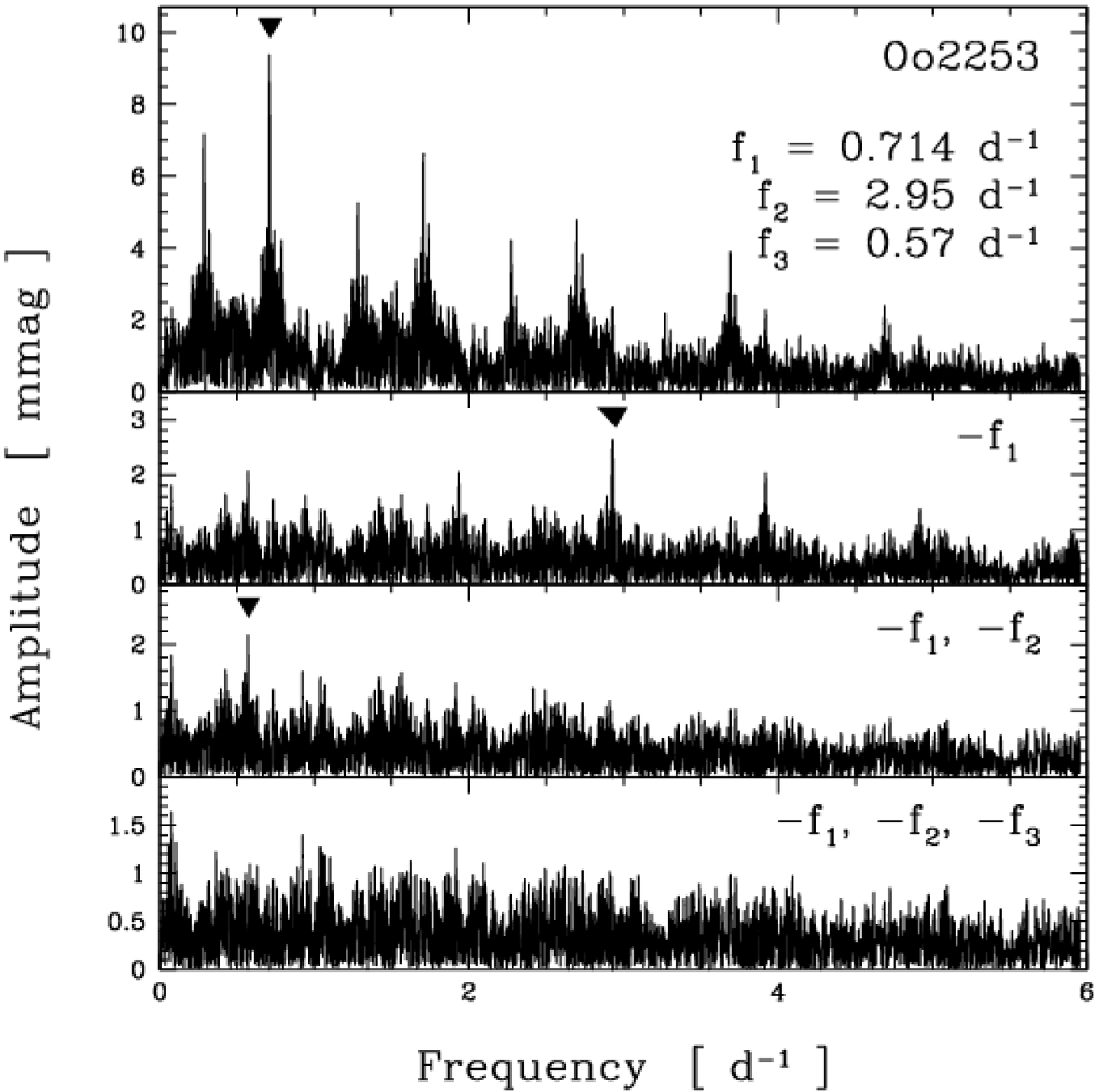}{Periodograms of subsequent stages
of prewhitening based on Polish and Chinese data of the
SPB star Oo~2253.}{spb}{!ht}{clip,angle=0,width=0.5\textwidth }
\figureCoAst{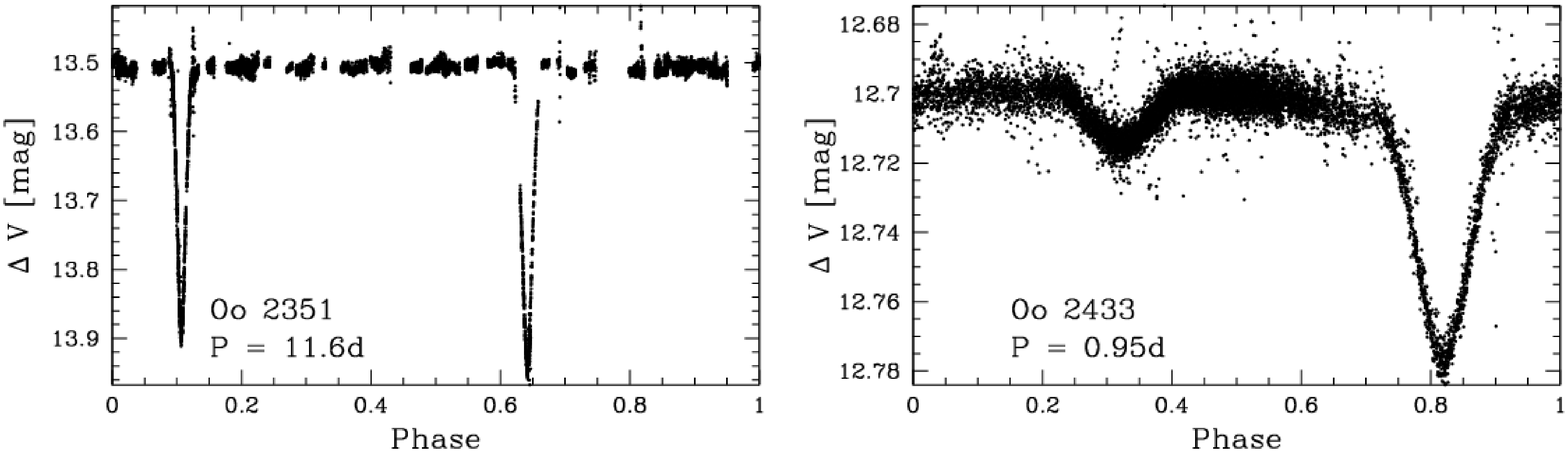}{Phase plots of two newly discovered eclipsing binaries
Oo~2351 and
Oo~2433.}{ecl}{!ht}{clip,angle=0,width=\textwidth }

\section{Future prospects}
First of all, we will conduct a detailed search for variable stars in the cluster
\ngc{884}. For these variable stars we will do a frequency analysis. We will
especially look for B-type pulsators, for which we will perform a mode
identification to determine the degree $\ell$. The well-known photometric
method, which compares the theoretical amplitude ratios with the observed ones
at different wavelengths, will be used (Dupret et al.~2003). For this purpose we
have observations in different bands at our disposal. The photometric amplitude ratios are
going to be combined with radial-velocity amplitudes deduced from simultaneous NOT spectra for the three brightest $\beta$ Cephei stars. This will make the mode
identification more conclusive than only the amplitude ratios (Daszy\'{n}ska-Daszkiewicz et al.~2005).

In the end we will fit theoretical frequencies to the observed ones and their
mode identification, simultaneously for all pulsating cluster members.
Indeed, as the stars in a cluster were born out of the same cloud, we
can assume that they have the same age and had the same chemical composition at
birth. The Li\`{e}ge stellar evolution code {\sc cl\'{e}s} (Scuflaire et
al.~2008) and the non-adiabatic oscillation code {\sc MAD} (Dupret et al.~2002)
will be applied in this process. Only models that fulfill additional
criteria, such as the position in the HR diagram derived from photometry, and the abundances of the stars obtained by NOT spectroscopy, will
be retained.

In any case, the first results are very promising for our future analysis of all campaign data. In-depth evaluation of stellar evolution models seems therefore within reach, now that the technique of asteroseismology has been extensively tested on single field stars.

\acknowledgments{
S.\,Saesen is an Aspirant Fellow and F.\,Carrier is a Postdoctoral Fellow of the
Fund for Scientific Research, Flanders (FWO). K.\,Uytterhoeven acknowledges financial support from a \emph{European Community Marie Curie Intra-European Fellowship}, contract number MEIF-CT-2006-024476.
}

\References{
\rfr Aerts, C., Thoul, A., Daszy\'{n}ska, J., et al. 2003, Science, 300, 1926
\rfr Ausseloos, M., Scuflaire, R., Thoul, A., \& Aerts, C. 2004, \mnras, 355,
352 
\rfr Ausseloos, M. 2005, dissertation, K.U.Leuven, Belgium
\rfr Daszy\'{n}ska-Daszkiewicz, J., Dziembowski, W. A., \& Pamyatnykh, A. A
. 2005, \aap, 441, 641
\rfr Dupret, M.-A., De Ridder, J., De Cat, P., et al. 2003, \aap, 398, 677
\rfr Dupret, M.-A., De Ridder, J., Neuforge, C., et al. 2002, \aap, 385, 563
\rfr Dziembowski, W. A., \& Pamyatnykh, A. A. 2008, \mnras, 385, 2061
\rfr Handler, G., Jerzykiewicz, M., Rodr\'{i}guez, E., et al. 2006, \mnras, 365,
327
\rfr Krzesi\'{n}ski, J., \& Pigulski, A. 1997, \aap, 325, 987
\rfr Krzesi\'{n}ski, J., \& Pigulski, A. 2000, ASPC, 203, 496
\rfr Pamyatnykh, A. A., Handler, G., \& Dziembowski, W. A. 2004, \mnras, 350,
1022
\rfr Pigulski, A., Handler, G., Michalska, G., et al. 2007, CoAst, 150, 191
\rfr Saesen, S., Pigulski, A., Carrier, F., et al. 2008, JPhCS, in press
\rfr Scuflaire, R., Th\'eado, S., Montalb\'an, J., et al. 2008, Ap\&SS, 316, 83
\rfr Stetson, P. B. 1987, PASP, 99, 191
\rfr Tamuz, O., Mazeh, T., \& Zucker, S. 2005, \mnras, 356, 1466
\rfr Waelkens, C., Lampens, P., Heynderickx, D., et al. 1990, A\&AS, 83, 11
}

\end{document}